\documentclass{iau} 
\usepackage{graphicx}

\title[IAU 340~~Long-Term Datasets for the Understanding of Solar and 
Stellar Magnetic Cycles]
{Building a Large Solar Analog Sample Using K2}

\author[Derek L. Buzasi]   
{Derek L. Buzasi$^1$}

\affiliation{$^1$Dept. of Chemistry \& Physics, Florida Gulf Coast University,\\ 10501 FGCU Blvd. S., Fort Myers, FL 33965 USA \\email: {\tt dbuzasi@fgcu.edu}}

\pubyear{2018}
\volume{340}  
\jname{Long-Term Datasets for the Understanding of Solar and Stellar Magnetic Cycles}
\editors{D. Banerjee, J. Jiang, K. Kusano, \& S. Solanki, eds.}
\begin{document}

\maketitle

\begin{abstract}
We have begun a project aimed at
providing a large consistent set of well-
vetted solar analogs in order to
address questions of stellar rotation,
activity, dynamos, and gyrochronology.
We make use of the K2 mission fields
to obtain precise photometric time
series, supplemented by ground-based
photometric and spectroscopic data for
promising candidates. From this data
we will derive rotation periods, spot
coverages, and flare rates for a well-
defined and well-calibrated sample of
solar analogs.
\end{abstract}

\firstsection 
\section{Sample Selection}

Our K2 sample is selected based on colors, for consistency with our earlier
Kepler sample (Buzasi, Lezcano, \& Preston 2015); we adopt $0.639 < B - V < 0.716$.
We have eliminated stars with known non-main sequence spectral types based
on catalog searches in SIMBAD and VIZIER, and used TRILEGAL simulations to
estimate that $<10\% $ of the remaining stars in our sample have $\log g<3.5$. Since
the light curves and power spectra can be used after the fact to eliminate
giants from the sample (Mathur {\it et al.} 2011), we view this level of contamination
as acceptable. We are focusing our attention on targets bright enough ($Kp<10$)
for ground-based spectroscopic follow-up with meter-class telescopes.
In support of this project, we have 11 approved K2 campaigns through
Campaign 17, supplemented by additional targets observed by other
investigators that fit our criteria. To date, data on approximately 2000 targets
have been released and analysis is under way.

K2 data are available through a number of different data
reduction pipelines, and outputs are not
necessarily consistent between pipelines. We compared results
from 6 different publicly available algorithms, including K2SC and K2SC-P (Aigrain,
Parviainen, \& Pope 2016), K2SP (Buzasi {\it et al.} 2016), EVEREST (Luger {\it et al}
2016, K2SFF (Vanderburg \& Johnson 2014), and the project pipeline product (K2; Van Cleve
{\it et al.} 2016).
Differences between output periods are frequently substantial,
particularly at low frequencies, and these
differences make the typical methodology used for photometric
range (range between 5th and 95th percentile) potentially
problematic.

\section{Period Detection, Comparisons, and Conclusions}
We analyzed output from each pipeline using four different period detection algorithms:
\begin{enumerate}
\item Lomb-Scargle Periodogram (LS): A commonly-used algorithm for detecting and
characterizing periodicity in unevenly-sampled time-series
\item Autocorrelation function (ACF): A measure of the correlation between values of the
time series at different times, as a function of the time lag.
\item Wavelet transform (Wave): Decomposition into families of basis functions that are
localized in both real and Fourier space.
\item Hilbert-Huang transform (HHT): Decomposition into empirical modes, followed by
application of the Hilbert transform to determine frequencies and amplitudes.
\end{enumerate}

In some cases, there is general agreement between the different approaches.
However, much of the time agreement is not so dependable, and this is particularly the
case for the long-period variability that is most of interest for rotating solar-like stars.

\begin{figure}[b]
\begin{center}
 \includegraphics[width=3.4in]{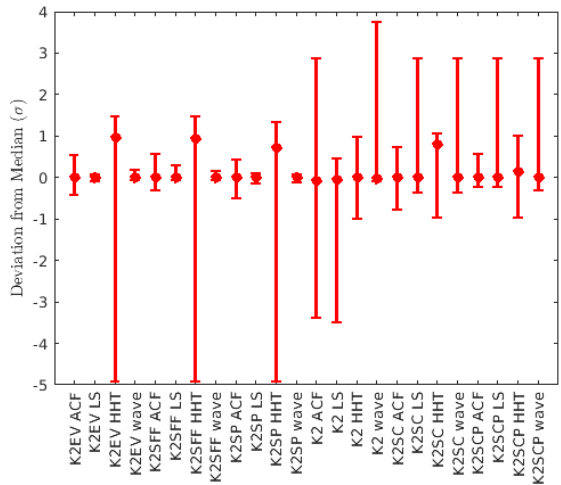} 
 \caption{Comparison of the performance of the different pipeline + period algorithm combinations
relative to the median value for each target. Deviations are measured in standard deviations.}
\end{center}
\end{figure}

Our 6 pipelines and 4 algorithms produce 24 potentially
different periods for each target, and we have compared each
of these period sets with
the ensemble median for each star (Figure~1). Systematic
effects due to some algorithms (ACF, HHT) are apparent, and it
is clear that some combinations (K2EV+LS, K2SP+Wavelet,
K2SP+LS, K2SFF+Wavelet, K2EV+Wavelet, K2SFF+LS) are
systematically more repeatable.

We are currently examining the most problematic algorithms to see if their
performance can be improved, as well as performing similar comparisons
for photometric range characterization. Once those comparisons are
complete, we will be able to make a well-informed decision regarding the
appropriate choice of tools for light curve characterization. This in turn will
allow us to efficiently direct our ground-based follow-up resources with the
goal of building a well-characterized sample of solar analogs.


\begin{thebibliography}{}


\bibitem[Aigrain, Parviainen, \& Pope (2016)]{Aigrain2016}
{Aigrain, S., Parviainen, H., \& Pope, B.} 2016,
\textit{MNRAS}, 459, 2408. (K2SC, K2SCP) 

\bibitem[Buzasi, Lezcano, \& Preston (2016)]{Buzasi2016}
{Buzasi, D., Lezcano, A., \& Preston, H.} 2016, 
\textit{JSWSC}, 6, 38. 

\bibitem[Buzasi et al. (2016)]{Buzasi_etal2016}
{Buzasi, D., Carboneau, L., Hessler, C., Lezcano, A., \& Preston, H. } 2016, 
\textit{Proc. IAU}, 29, 673. (K2SP)

\bibitem[Luger et al. (2016)]{Luger2016}
{Luger, R. {\it et al}} 2016, 
\textit{AJ}, 152, 100. (K2EV)

\bibitem[Mathur et al. (2011)]{Mathur2011}
{Mathur, S. {\it et al.}} 2011, 
\textit{ApJ}, 741, 119.

\bibitem[Vanderburg \& Johnson (2014)]{Vanderburg2014}
{Vanderburg, A. \& Johnson, J.} 2014, 
\textit{PASP}, 126, 948. (K2SFF)

\bibitem[Van Cleve (2016)]{VanCleeve2016}
{Van Cleve {\it et al.}} 2016,
\textit{PASP}, 128, 5002. (K2)

\end{thebibliography}
\end{document}